\documentclass[manuscript]{aastex}	
\usepackage{epsfig}			

\usepackage{graphicx,color}		
\usepackage{amssymb}			
\usepackage{color}			
\usepackage{url}			
\usepackage{amsmath}			
\usepackage{rotating}			
\usepackage{float}			
\usepackage{textcomp}			
\usepackage{psfig}
\usepackage{dcolumn}
\usepackage{times}
\usepackage{tabularx}
\usepackage[colorlinks=true,citecolor=blue]{hyperref}
\usepackage[english]{babel}

\begin{document}
\title{Long term study of the solar filaments from the Synoptic Maps as derived from $H_{\alpha}$ Spectroheliograms of Kodaikanal Observatory}
\author{Subhamoy Chatterjee$^{1}$,
Manjunath Hegde$^{1}$,
Dipankar Banerjee$^{1,2}$,
B. Ravindra$^{1}$,
}
  \affil{$^{1}$Indian Institute of Astrophysics, Koramangala, Bangalore 560034, India. e-mail: {\color{blue}{dipu@iiap.res.in}}\\
$^{2}$Center of Excellence in Space Sciences India, IISER Kolkata, Mohanpur 741246, West Bengal, India  \\}

\begin{abstract}
The century long (1914-2007) $H_{\alpha}$ (656.28 nm) spectroheliograms from Kodaikanal Solar Observatory (KSO) have been recently digitised.  Using these newly calibrated, processed images we study  the evolution of dark elongated on disk structures called filaments, potential representatives of magnetic activities on the Sun. To our knowledge this is the  oldest uniform digitised dataset with daily images available today in $H_{\alpha}$. We generate Carrington maps for entire time duration and try to find the correspondences with maps of same rotation from Ca II K KSO data. Filaments are segmented from Carrington maps using a semi-automated technique and are studied individually to extract their centroids and tilts. We plot the time-latitude distribution of filament centroids producing  Butterfly diagram, which clearly shows presence of poleward migration. We separate polar filaments for each cycle and try to estimate the delay between the polar filament number cycle and sunspot number cycle peaks. We correlate this delay with the same between polar reversal and sunspot number maxima. This provides new insight on the role of  polar filaments on polar reversal. 
\end{abstract}

\keywords{Sun: chromosphere --- Sun: filaments ---  techniques: image processing --- methods: data analysis}
\section{Introduction}
Sun is a star which is highly heterogeneous both temporally and spatially. It is crucial to study the  long term evolution of the magnetic field, which is believed to be the major driver for this heterogeneity.  Before 1970s regular magnetic field measurements were not available and thus  long term study of proxies of solar magnetic field has been an important subject area.   Solar filament has served as one such proxy \citep{1972RvGSP..10..837M} for understanding the magnetic activity of the Sun. They are clouds of ionized gas projected against the solar disk 
which are cooler and denser than the plasma underneath. In $H_{\alpha}$ ($\approx 656.28$ nm line center)
observations they appear as dark, elongated, thin hairy structures. 
Filaments are formed along the polarity inversion line between opposite magnetic polarities \citep{Martin1998}. When filament
becomes unstable due to magnetohydrodynamic instability, it erupts.
Filament eruption is often associated with flare and coronal mass
ejections (CMEs), which are the sources of geomagnetic storms
 \citep{{2000AdSpR..25.1851G},{0004-637X-537-1-503},{0004-637X-586-1-562},{2008A&A...484..487C},{2012A&A...542A..52Z}}.

 Solar filaments are seen at all latitudes on the solar disk from 
equator to pole during all phases of the solar cycle and outline the boundary between different
magnetic field polarities. This makes them ideal candidate for the study of 
large scale concentrations  of  weaker magnetic field on the sun \citep{{1972RvGSP..10..837M},{ROG:ROG841},{1983SoPh...85..227M}}. Thus, long term filament data can be efficiently used as a proxy for magnetic activity of the sun. Additionally, study of the occurrence of filaments give useful insight to distribution of these fields on the solar surface, 
their evolution and to understand the nature of sun's magnetic field 
\citep{1994A&A...290..279M}. Detection of filaments from the archival images is the first step for such a study. There have been many attempts to detect filaments in automated way from full disk $H_{\alpha}$ images. The methods range from modified Otsu thresholding \citep{2015ApJS..221...33H} to application of artificial neural networks \citep{Zharkova2003}. But, most of these methods were applied only on good quality digital images of few recent solar cycles. Historical digitized data present itself with many inherent artefacts which can not  be corrected by flat fielding \citep{zhar}. Application of most automated methods on such data mostly result in under-estimation or over-detection of filaments. So, a careful detection of such features are necessary and it may require manual intervention. Detection of filaments from Carrington/synoptic maps of historical data have generated promising results as reported by few recent studies. \citet{Li2007} used synoptic 
charts of filament archive from Meudon observatory to study long term variation of solar filaments. 
Using the same data set, \citet{doi:10.1111/j.1365-2966.2010.16508.x} studied latitude migration of filaments at 
low latitudes (less than $50^\circ$) and found latitudinal drift of filaments differs from 
that of sunspot groups.  \citet{2015ApJS..221...33H} utilized Big Bear Solar Observatory (BBSO) data from 1988 -- 2013 to 
extract variation of filament area, spine length, tilt angle and barb number with respect to 
calendar year and latitude. The study also included North -- South asymmetry of filament number. 
\citet{{2016ASPC..504..241T},{2016SoPh..291.1115T}} using synoptic charts from Meudon observatory and Kislovodsk Mountain 
Astronomical station studied variation of filament tilt angle and classified  filaments with 
different tilt angle viz., active region filaments, quiescent filaments and polar filaments. There have been some works from Kodaikanal historical data on prominences \citep{{1952Natur.170..156A},{1907MNRAS..67..477E},{1908MNRAS..68..515E},{1993SoPh..146..259R}}, flares \citep{{1993SoPh..146..137R},{1994SoPh..149..119S},{2001BASI...29...77S}}. Most of these earlier studies were for shorter period of time.  \citet{1952Natur.170..156A} used Ca {\sc ii} K disk  blocked plates to study the evolution of prominences till 1954 and \citet{1983SoPh...85..227M} used kodaikanal Ca {\sc ii} K and $H_{\alpha}$ plates till 1919 to study poleward migration of magnetic neutral lines.
In this paper we present for the first time, the newly digitized unique data set of $H_{\alpha}$ filaments from KSO and study 
solar activity variation for 9 solar cycles. To our knowledge this is the oldest digitised archive of $H_{\alpha}$. 
In section 2, we present the description of the data. Method of analysis, and representative results are presented in sections 3 and 4 respectively. Finally in section 5 we summarise our results and  discuss the potential of such unique archive for future studies. 
\begin{figure}[!thb]
\centerline{
\includegraphics[scale=0.9]{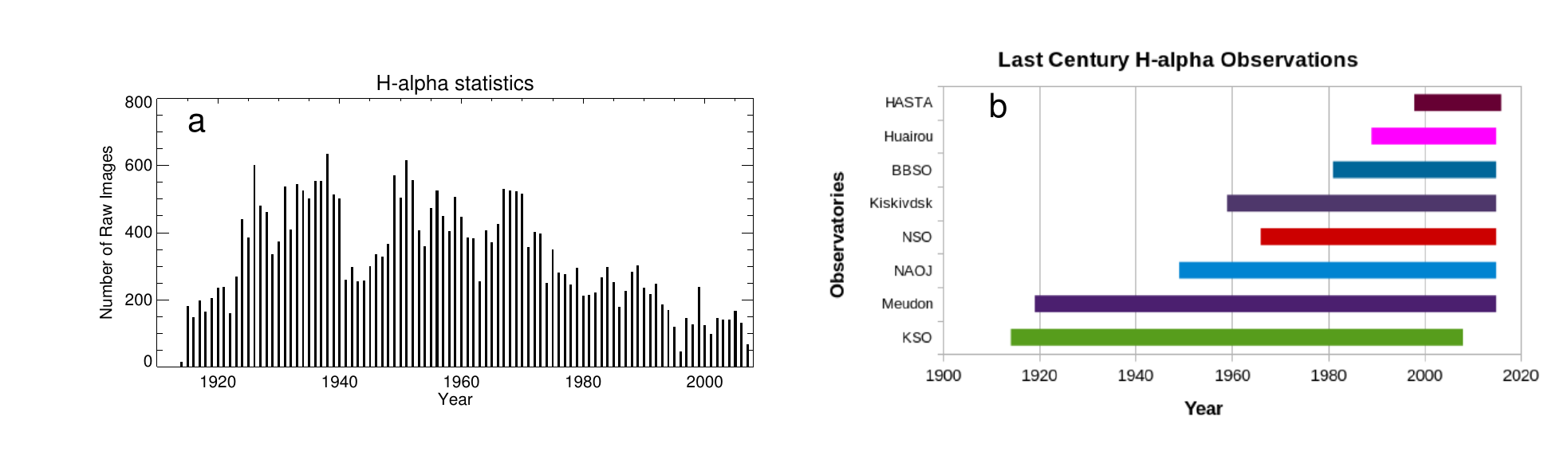} }\vspace{-0.5cm}
\caption{Availability of $H_{\alpha}$ images. a) Yearly histogram of KSO $H_{\alpha}$ data; b) Availability of $H_{\alpha}$ dataset across the globe. Different sources are marked with abbreviations along y-axis. }

\label{histogram}
\end{figure}
\section{Data Description}
Systematic $H_{\alpha}$ observations were carried out at Kodaikanal Solar Observatory (KSO), India since 1914  with a telescope (spectroheliograph) having  30 cm objective and f-ratio of 21. The spectroheliograms (656.3~nm) were consistently recorded in photographic plates until 1978 and subsequently in films on daily basis. These plates/films have recently been digitised by the help of a uniform illumination source and a 4096 $\times$ 4096 CCD cooled at -100$^\circ$C. We used those digitized full disk 4096 $\times$ 4096 KSO $H_{\alpha}$ density images from 1914 to 2007 ($\approx$ 0.86 arcsec /pixel) in our current study. The seeing limited resolution has been $\approx $ 2 arcsec for majority of observing period. Figure ~\ref{histogram}a shows the number of images available in the archive (used in this paper) and Figure ~\ref{histogram}b shows the time span over which the KSO digitized images are available along with other major data sources.   Observation for time duration close to Kodaikanal has been by the Meudon Observatory, France (since 1919) (Figure ~\ref{histogram}b). Recent studies with this data has been performed by  \citet{{2016ASPC..504..241T},{2016SoPh..291.1115T}} combining  $H_{\alpha}$ observation of Kislovodsk  Mountain Astronomical Station, Pulkovo Astronomical Observatory, the Russian Academy of Sciences after 1959.\\

\begin{figure}[!thbp]
\centerline{
\includegraphics[scale=0.55]{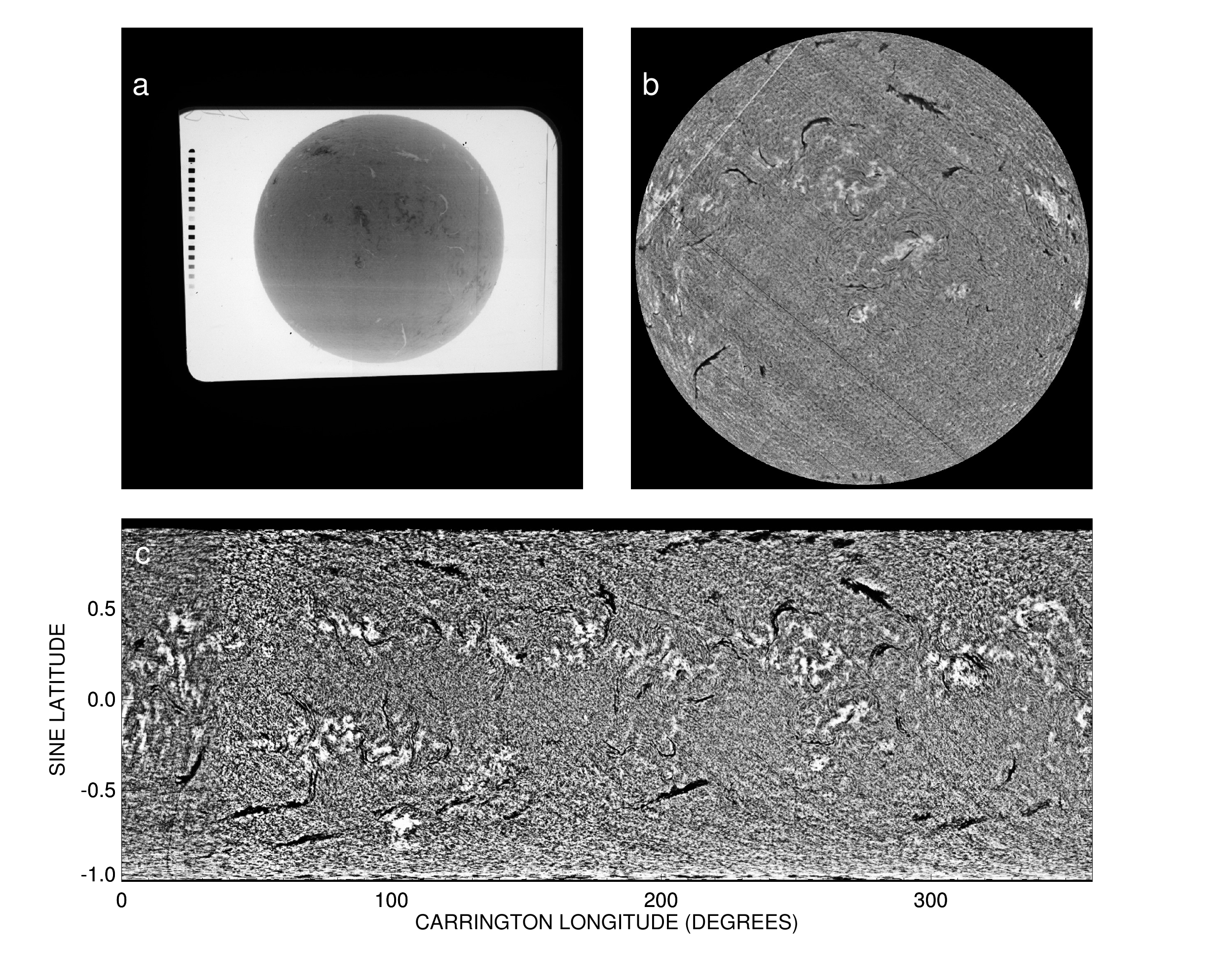}}
\caption{KSO $H_{\alpha}$ data calibration and filament detection. a) $H_{\alpha}$ RAW image taken on March 16, 1981; b) Limb darkening corrected and disk centered image; c) Carrington map for rotation number 1706 starting on March 8, 1981.}
\label{processing}
\end{figure}

\begin{figure}[!thbp]
\centerline{
\includegraphics[scale=0.8]{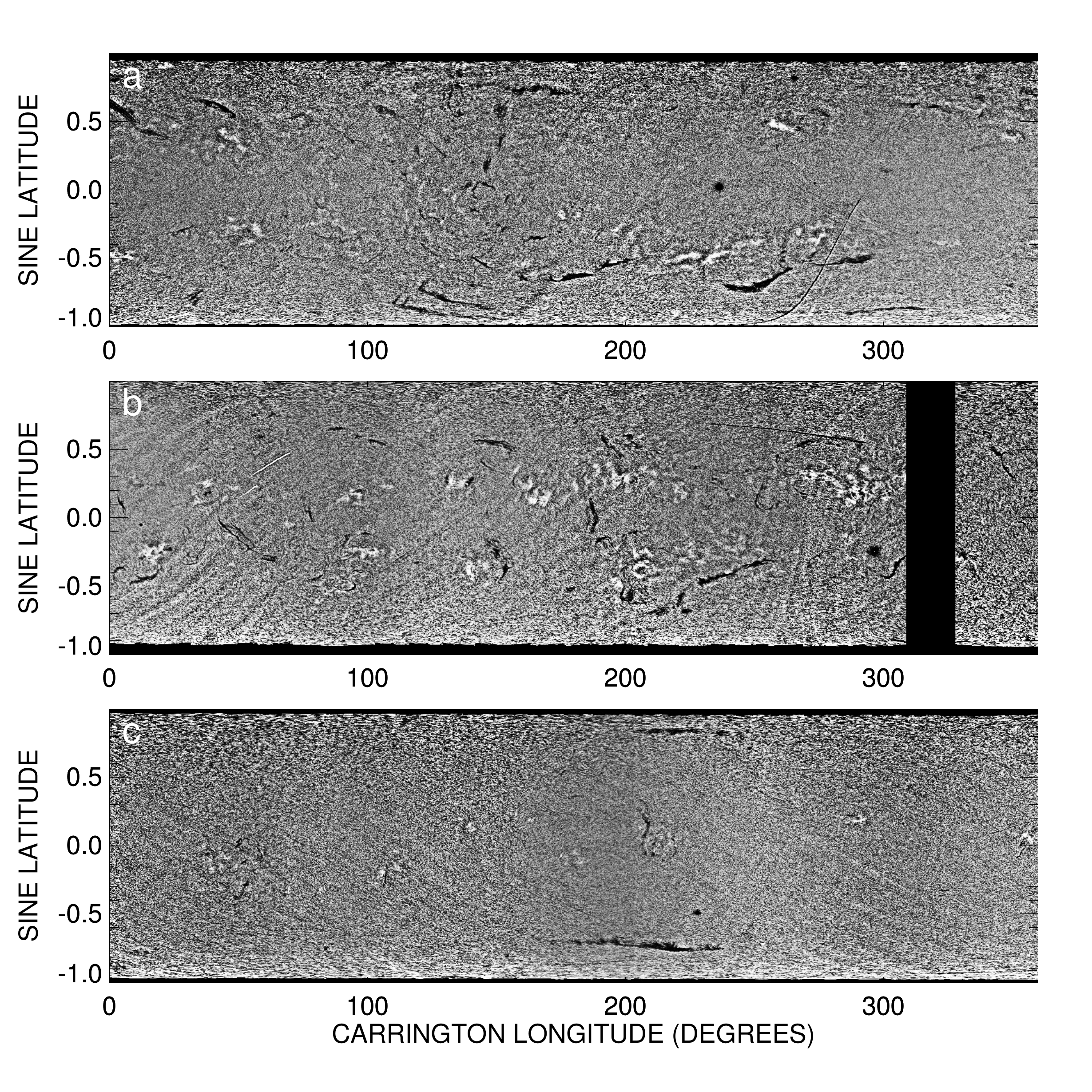}}
\caption{KSO $H_{\alpha}$ maps for different phases of solar cycles. a) Carrington rotation 970 depicting a representative filament distribution at the rising phase of cycle 16; b) Carrington rotation 1001 depicting a representative filament distribution at the maxima of cycle 16; c) Carrington rotation 1641 depicting a representative filament distribution at the minima of cycle 20.}
\label{3maps}
\end{figure}
\section{Methods}
\subsection{Calibration}
The calibration includes several steps. First  the RAW images (Figure ~\ref{processing}a) were inverted in grey scale. Edge detection operators were applied to produce a binary edge detected image. Circle Hough transform \citep{sonka2014image} was applied on that binary image to find the center and radius of disk. After disk centering the images were resized to a smaller version and were median filtered to obtain the asymmetric limb darkening profile combining effects of  line of sight and imaging instrument. This step was applied to reduce the time complexity of the automated calibration per image as the limb darkening profile  captures only the large scale intensity variation. The profile was again blown up to original size and original image was divided with the same to generate limb darkening corrected images as shown in Figure ~\ref{processing}b.  
\subsection{Carrington map generation}
Carrington map is a type of Mercator projection of the spherical sun generated from synoptic observation corresponding to one solar rotation. In this study, 60$^\circ$ longitude bands (-30$^\circ$ to 30$^\circ$ heliographic longitude) in the limb darkening corrected full disk $H_\alpha$  images were selected. Those were $B_0$ angle corrected and mapped to the Carrington longitude-latitude grid  in form of a rectangle with a weightage of cosine 4th power \citep{2011ApJ...730...51S} over each heliographic longitude. These rectangular slices were shifted and added according to date and time for 27.2753 days to generate a full 360 degree map of sun. A similar 360 degree map was obtained from rectangular binary slices called streak map \citep{{2011ApJ...730...51S}, {0004-637X-827-1-87}}. The overlap of same Carrington longitudes was removed through division of the original solar 360 degree map with streak map to form an image called Carrington map or Synoptic map. One representative Carrington map is shown in Figure ~\ref{processing}c. A total of 1185 Carrington maps from rotation 817 to 2062 are posted on this portal (\url{ftp://ftp.iiap.res.in/subhamoy/halpha_carrington_maps_kodaikanal/}). Figure ~\ref{3maps} depicts three representative Carrington maps for three different phases of solar cycles. Figures ~\ref{3maps}a, \ref{3maps}b and \ref{3maps}c present respectively rising phase, maxima of cycle 16 and minima of cycle 20. It can be observed from the maps that polar filaments are dominantly present during rising phase and minima of the solar cycles whereas low latitudes filaments dominantly occur in-between i.e. around cycle maxima. It should be noted that bright and dark ridge like structures in the Carrington maps are not of solar origin. Scratch lines present in the RAW images are responsible for those and they manifest as curved shape in the carrington maps due to north-south rotation correction. 
\subsection{Filament detection and parameter extraction}
\begin{figure}[!thbp]
\centerline{
\includegraphics[scale=0.8]{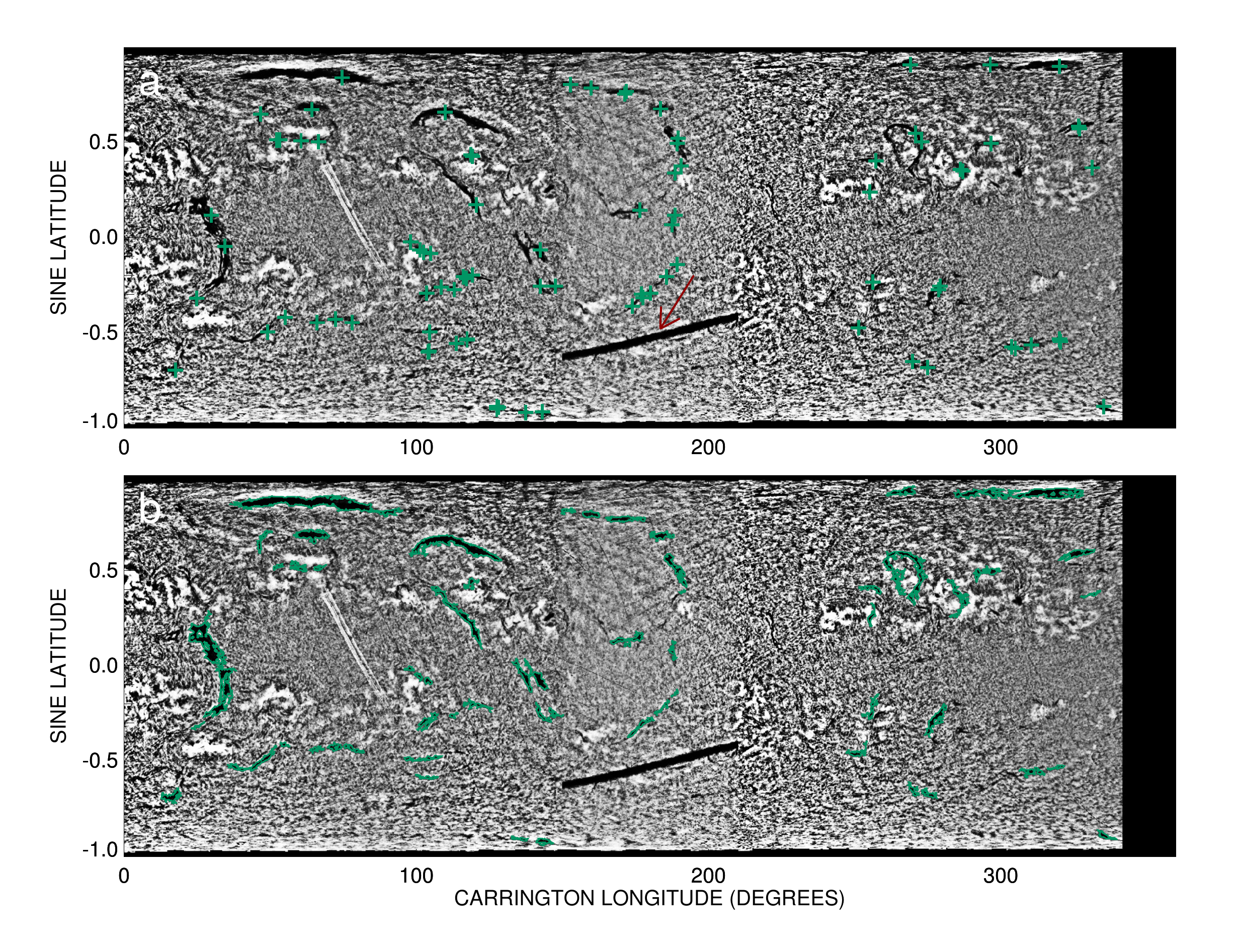}}
\caption{Region growing for KSO $H_{\alpha}$ filament detection. a) $H_{\alpha}$ Carrington rotation 1823 with seed points marked with green `+' symbols. Scratch line resembling a filament is marked with red arrow; b) Green contours depict filament boundaries detected through region growing about the manually selected contours.}
\label{region}
\end{figure}

\begin{figure}[!thbp]
\centerline{
\includegraphics[scale=0.7]{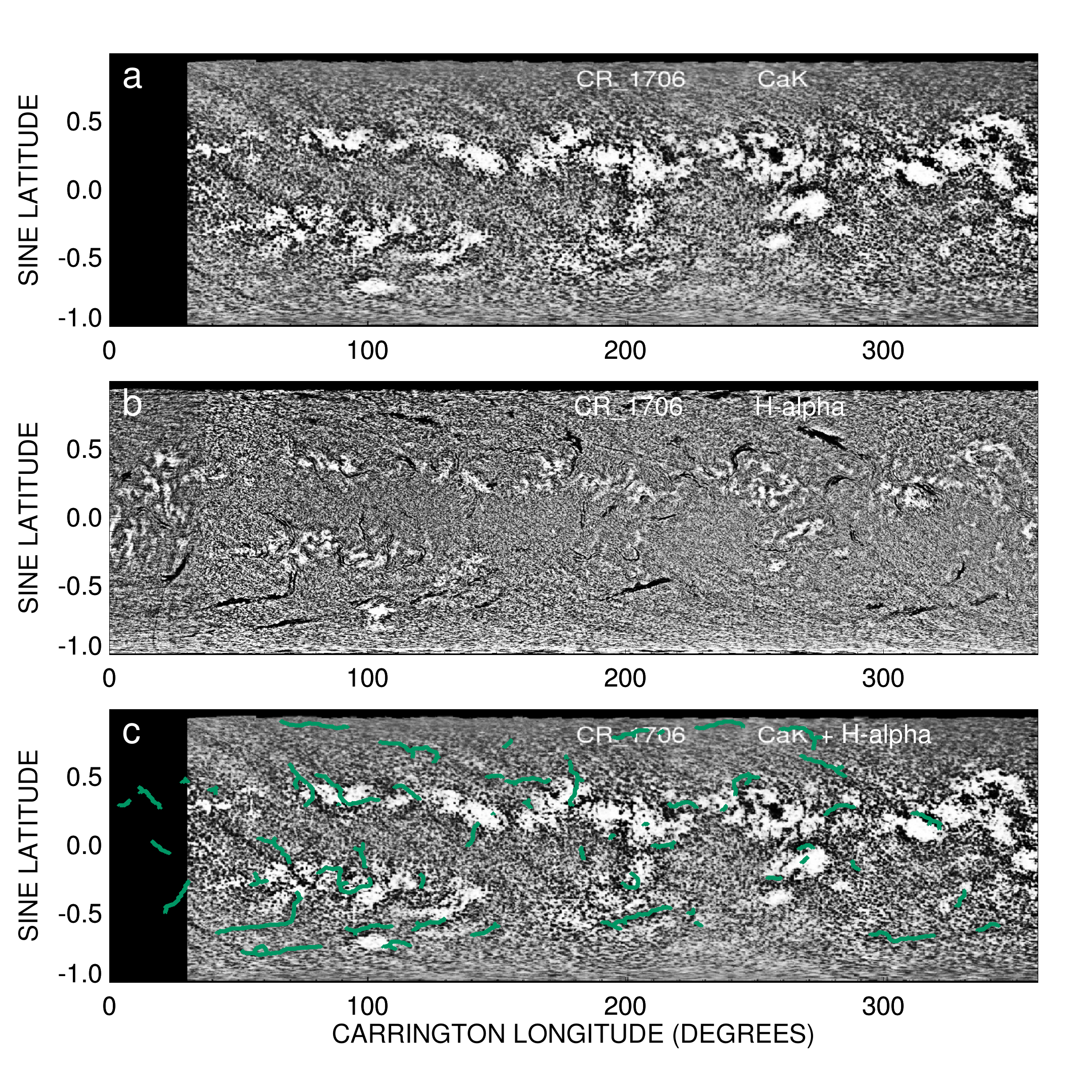}}
\caption{Correspondence between KSO Ca {\sc ii}  K and $H_{\alpha}$ maps. a) KSO Ca {\sc ii} K Carrington rotation 1706; b) KSO  $H_{\alpha}$ Carrington rotation 1706; c) Filament spine in green from (b) over plotted on (a).}
\label{cak_hal}
\end{figure}

\begin{figure}
  \vspace{-.08\textwidth}
  \centering
  \includegraphics[width=0.8\linewidth]{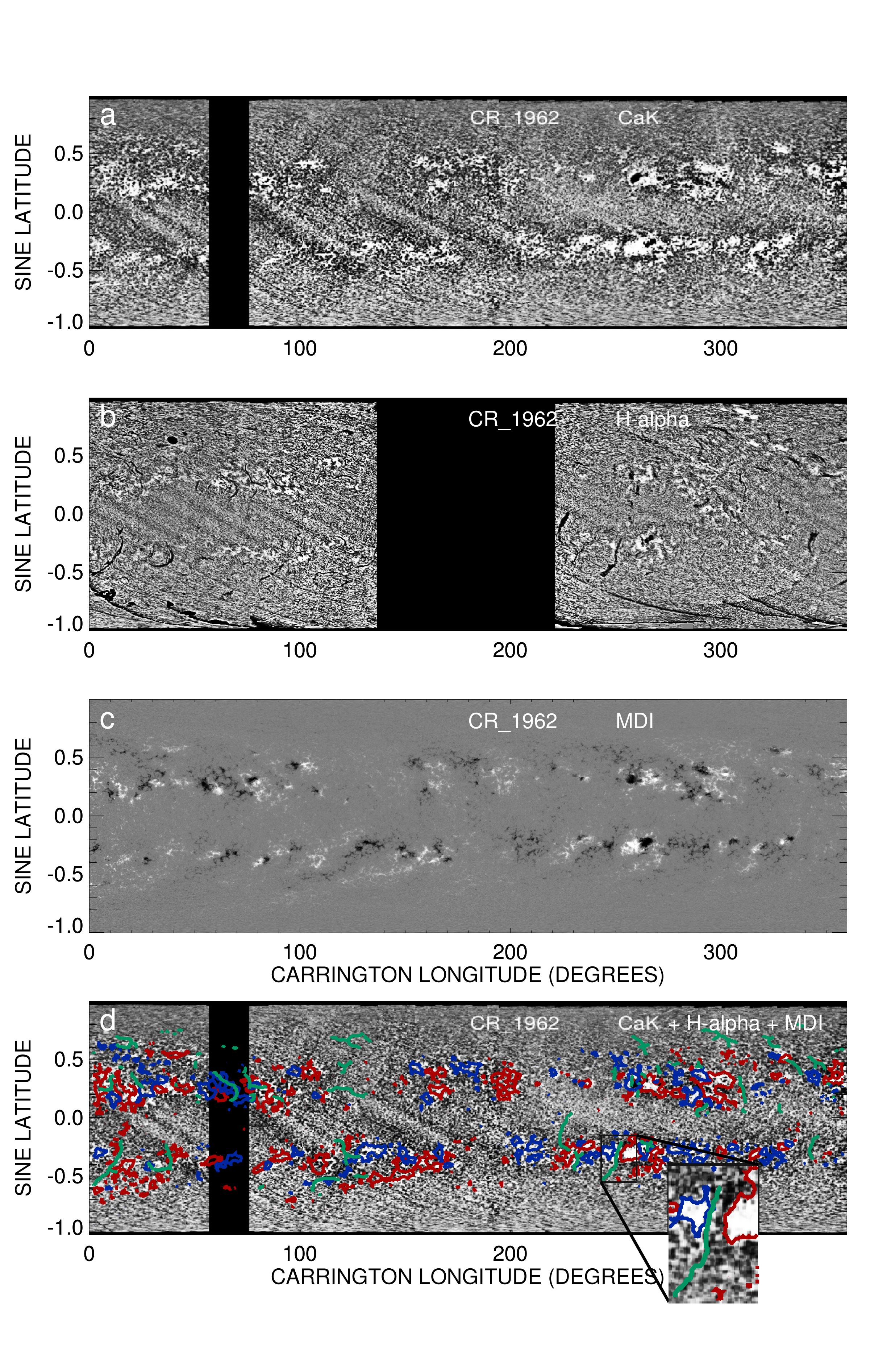}
    \vspace{-.055\textwidth}
  \caption{Correspondence between KSO Ca {\sc ii}  K, $H_{\alpha}$ and MDI magnetogram maps. a) KSO Ca {\sc ii}  K Carrington rotation 1962; b) KSO  $H_{\alpha}$ Carrington rotation 1706; c) Carrington map generated from MDI/SoHO full disc LOS magnetograms for rotation 1962; d) Filament spine in green from (b) over plotted on (a) with positive (red) and negative (blue) magnetic field contours of rotation 1962 from MDI/SoHO. This panel also shows a magnified view of small window within the Carrington map depicting a filament lying in between two magnetic patches of opposite polarity.}
 \label{fig:cak_hal_mdi}
\end{figure}

Figure ~\ref{region}a shows the Carrington rotation 1823 with many filaments and an artefact mimicking a filament marked with green arrow. If a fully automated filament detection algorithm is used the artefact will be detected as filament. This was the reason we went for semi-automated  technique. 
The Carrington maps were first intensity enhanced through histogram equalization. Seed points (shown with `+'  in Figure ~\ref{region}) were selected manually on the dark filaments. 8-neighbourhood region growing \citep{sonka2014image} was performed to detect filament for each selected seed point. The region grows by including neighbour pixels about the seed point satisfying intensity threshold. This process is repeated for each new neighbour pixel until pixels go totally outside intensity threshold and a connected region is produced. As we see in Figure ~\ref{region}a some filaments need multiple seed point selection for sudden jumps of intensity above the threshold selected for region growing. Subsequently, binary filament detected Carrington maps are generated. Contour of the produced binary map for rotation 1823 is overplotted in green on the gray scale Carrington map in Figure ~\ref{region}b. Because of image contrast, data gaps and temporal evolution naturally Carrington maps depict unnatural fragmentation in filaments. As the polar filament (latitude $>50^\circ$) are longer, their lengths are affected more. Morphological closing operations with different kernel sizes are performed for higher latitude filaments to examine change in number of polar filaments. Parameters such as centroid longitude,  latitude and filament tilts are generated. In Figure ~\ref{cak_hal}a we show the  Ca {\sc ii} K carrington map as generated from the KSO data to compare the  correspondence between the bright plage locations in Ca {\sc ii} K (Figure ~\ref{cak_hal}a) and  their traces in $H_{\alpha}$ (Figure ~\ref{cak_hal}b) corresponding to the rotation 1706. To make correspondences between filaments and active regions more evident, we plotted the filament skeletons on the Ca {\sc ii}  K as shown in Figure ~\ref{cak_hal}c. Using Michelson Doppler Imager (MDI) on-board SoHO line-of-sight magnetograms available after the year 1996 we generated Carrington maps and Figure ~\ref{fig:cak_hal_mdi} includes one such MDI map for inspecting the location of filaments in terms of distribution of magnetic fields. Carrington rotation 1962 has been compared from Ca {\sc ii}  K, $H_{\alpha}$ and MDI in Figures ~\ref{fig:cak_hal_mdi}a, ~\ref{fig:cak_hal_mdi}b and ~\ref{fig:cak_hal_mdi}c respectively. Formation of filaments along magnetic neutral lines can be observed at different locations and one such example is shown in  the zoomed inset of  Figure~\ref{fig:cak_hal_mdi}~d.  

\pagebreak
\section{Results}
\begin{figure}[!t]
\centerline{
\includegraphics[scale=0.75]{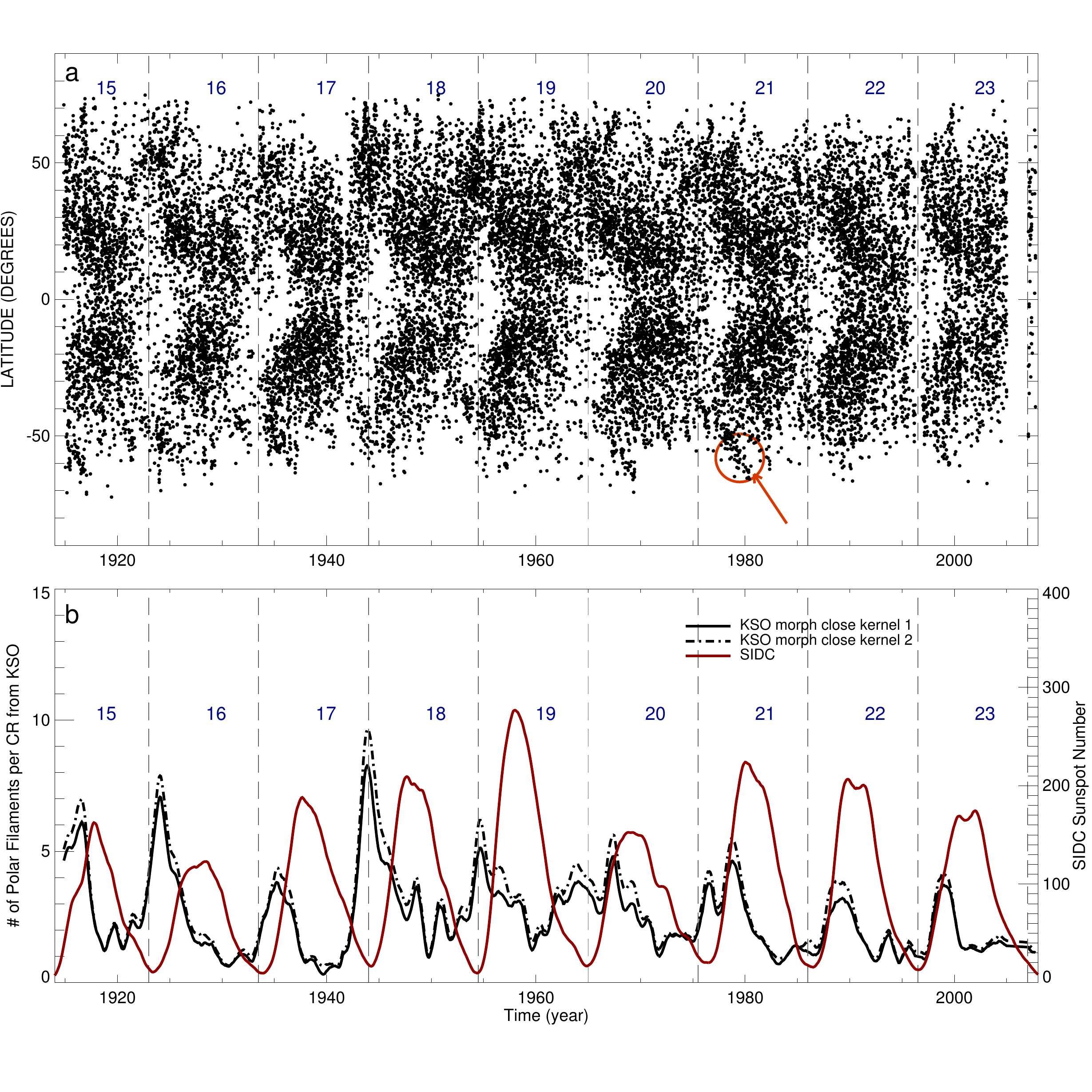} }
\caption{Distribution of filaments. a) Time-latitude distribution (1914 -- 2007) of solar filaments as detected from KSO Carrington maps. A representative signature of  poleward migration  is marked by a red circle and arrow. Cycle numbers are printed in blue with dashed vertical lines marking the cycle minima; b) Temporal variation (19014 -- 2007) of number of polar filaments as recorded from KSO and its comparison with SIDC sunspot number.  The solid black curve and dot-dash correspond to filaments morphologically closed by a disc kernel of radius 22 pixels and 30 pixels respectively. It is worth noting that temporal locations of filament number maxima do not change with different sizes of closing operation except for the relative change in number of filaments.}
\label{butterfly}
\end{figure}

Figure ~\ref{butterfly}a shows temporal evolution of the filament centroid latitudes over 9 cycles. This plot illustrates how the filaments are distributed at all latitudes
and also reveals (like sunspots) that  filaments also migrate towards equator (butterfly diagram) but from a higher latitude. 
Along with the butterfly diagram like nature, signature of poleward migration is observed in the plot. One such example is highlighted by a red circle with an arrow.
Corresponding to cycles 15, 17, 20, 21 and 23, in southern hemisphere rush to the pole is clearly seen. Similarly, in northern hemisphere, cycles 15,16, 20, 21 and 23 show this behaviour. Though there are traces, poleward rush is not very clear for cycles 18, 19, 22. The observed polar rush can be compared with the same from Ca {\sc ii} K prominence results as presented  in \citet{1952Natur.170..156A}. Though we don't observe the rush till pole because of projection effects on the on-disk feature, we find the early phases for most of the cycles. 
\begin{figure}[!t]
\centerline{
\includegraphics[scale=0.75]{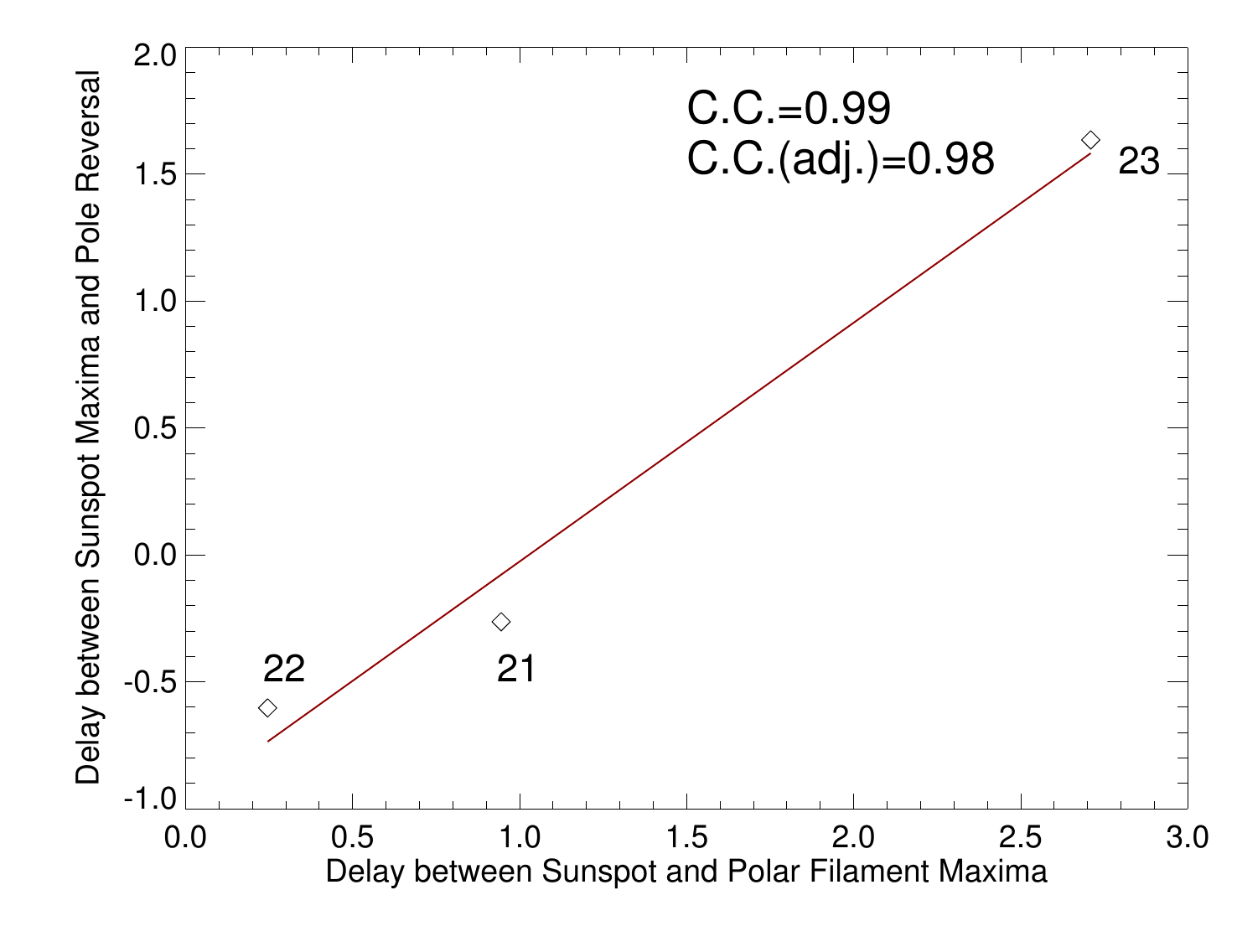} }
\caption{Correlation of the sunspot-polar filament number maxima delay and sunspot number maxima-pole reversal delay for last three cycles.  As number of data points is only 3, adjusted correlation coefficient is also shown. }
\label{pole_rev}
\end{figure}
\begin{figure}[!t]
 \centerline{
\includegraphics[scale=0.7]{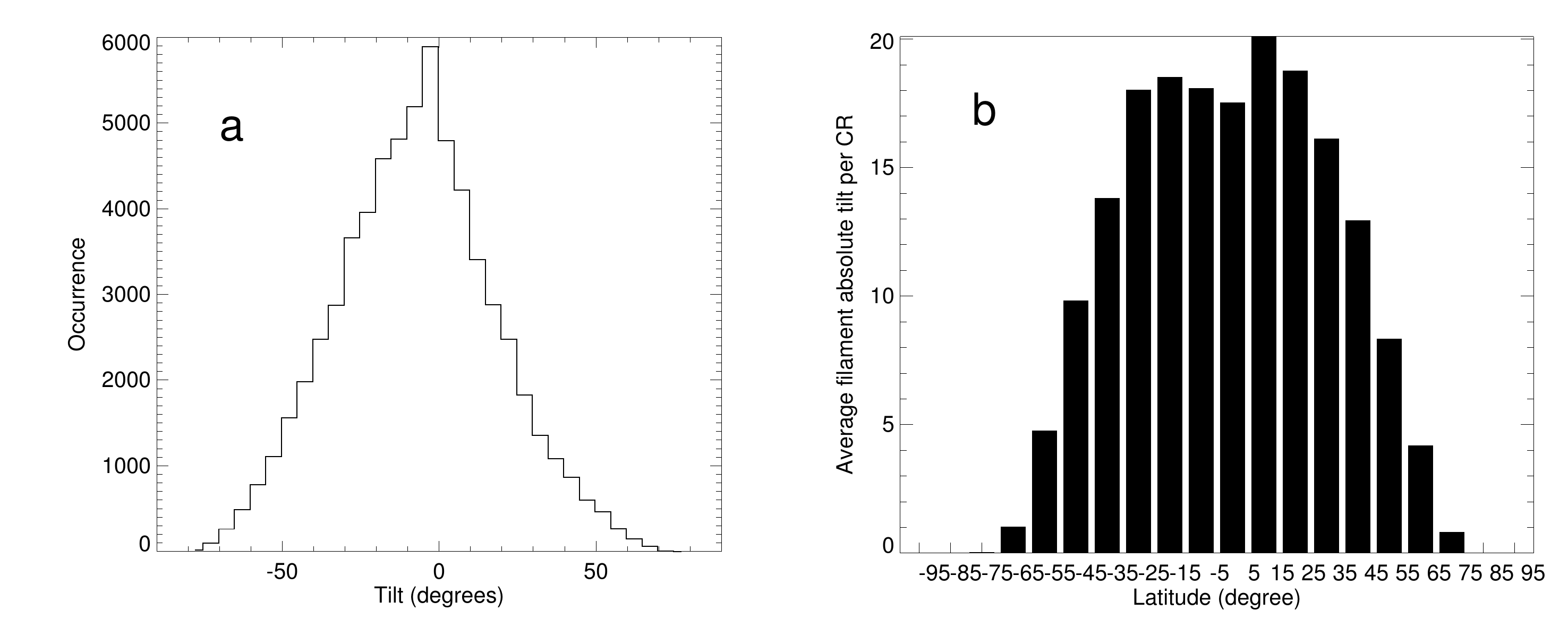} }
\caption{Filament tilt attributes. a) Histogram of filament tilt angle with respect to equator; b) Latitudinal distribution of solar filament tilts.}
\label{lentil}
\end{figure}
Now we focus our attention to the polar filaments.  Figure ~\ref{butterfly}b shows the temporal variation of the number of polar filaments (centroid latitude $> 50^\circ$) overplotted on smoothed sunspot  number (SSN) from Solar Influences Data Center (SIDC; \url{http://www.sidc.be/silso/datafiles}). We must point out here that  we find some cases of fragmentation in filaments during detection and at this point we don't have a method to estimate the correct length for such cases. Thus we concentrate on number density of filament rather than exact length estimates. We apply Morph close operations  on  the detected filaments with disc kernels of radii 22 and 30 pixels named as kernel 1 and kernel 2 respectively in Figure~\ref{butterfly}b. The target Carrington maps are of sizes $1570 (x)\times500 (y)$ with x-axis or longitude spanning from $0^\circ$ to $360^\circ$ and y-axis or sine latitude spanning over -1 to 1. In this exercise  we explore whether the temporal location of polar filament number maxima changes varying the kernel sizes. Figure~\ref{butterfly}b illustrates the natural reduction of filament number with higher kernel size though there is no apparent change in the epoch of maxima. Both the time series clearly present 11 year periodicity with same delay between their peaks and SSN maxima.  Based on  Wilcox Solar Observatory polar field data (taken from \url{http://wso.stanford.edu/}) we plot the SSN-polar filament maxima delay vs SSN maxima-pole reversal delay in Figure ~\ref{pole_rev} from three cycles namely 21, 22 and 23. For understanding the epoch of pole reversal (North-South)/2 curve is considered. They show a linear correlation ($r$) of 0.99. As the number of data points are limited, we also calculate the `Adjusted Pearson Correlation Coefficient' defined by the expression, $$r_{\rm adjusted}^2=1-(1-r^2)\frac{n-1}{n-p-1}$$ 
where, $n$ is the number of data points which is 3 in our case and $p$ is the number of independent variables which is 1 here.
We find the adjusted coefficient to be 0.98 which still remains pretty high confirming the association between epoch of number of polar filament peak and pole reversal with respect to sunspot maxima. 
This emphasises the role of polar filaments in polarity reversals. Faster the polar filaments migrate to poles, earlier will be the pole reversal. To our understanding this is a very important finding and demands a detailed study, which we plan to do in near future.  


Now we will look at the tilt angle distribution. Fig~\ref{lentil}a illustrates histogram of solar filament tilt angle with respect to equator. It shows an asymmetric nature consistent with the findings of \citet{2016SoPh..291.1115T}. We also take average of tilt angle absolute values for different latitude bands. This latitudinal distribution is plotted in Fig~\ref{lentil}b.
It is consistent with the Joy's law for sunspot pairs. Though we don't estimate the exact length of the filaments we believe that the average tilt of the fragmented filaments (a fraction of the total sample) are more or less the same as the fitted straight line slope of the whole filament and thus this plot should not change much due to filament fragmentation. As the filaments are expected to be oriented along neutral lines, they become more parallel to the equator at higher latitudes where the sunspot pairs (appearing till $\approx40^\circ$) become more tilted. It is worth mentioning that filaments appear allover the disk reaching much higher in latitudes compared to sunspot pairs and they maintain the trend of declining tilt with respect to equator. Thus the filament distribution may provide much detailed insight on the distribution of magnetic field at all latitudes rather than the sunspot locations alone. 
\section{Conclusions}
 In this paper, we prsent the calibration (no density to intensity calibration has been performed yet) and processing  of RAW $H_{\alpha}$ dataset (1914 -- 2007) from the Kodaikanal Solar Observatory. We have generated Carrington maps from the calibrated and processed daily spectroheliograms for the whole study period. A semi-automated algorithm relying on seed selection and region growing has been used to consistently detect filaments from the Carrington maps. From, the detected maps we have generated parameters such as filament centroid latitude and tilt angle with respect to the equator. Salient features of the new findings are listed here.

\begin{itemize} 
 \item We have generated the time-latitude distribution of filaments giving a clear signature of polar rush for several cycles along with butterfly diagram like pattern. It is worth noting that Cycle 23 shows a comparable signature of polar rush as that presented in Figure 5 of \citet{2015ApJS..221...33H}. 
 
\item  We segregated polar filaments with centroid latitude greater than $50^\circ$ and plotted their number for all the Carrington maps over time. The plot depicted similar 11 year periodicity as normally observed for sunspot number cycle with delays. We measured those delays between polar filament and sunspot number maxima. We could also get the epochs of polar magnetic field reversal from Wilcox Solar Observatory data for last three cycles. Those have have been used to correlate the sunspot number-polar filament delay and sunspot number-pole reversal delay. With limited  study points, we found out very high correlation indicating the role of polar filaments in polar reversal. This new result along with the rate of drift of filaments towards the pole may shed new light on the polar reversal.
  
  \item We found out asymmetry in the histogram of filament tilts consistent with the finding of \citet{2016SoPh..291.1115T}. Also, we could get the behaviour similar to Joy's law in the latitudinal distribution of filament tilts. It is believed that the tilt angle of sunspots and bipolar groups determines the conversion efficiency of the toroidal field into the poloidal magnetic field \citet{2016SoPh..291.1115T}. As filaments lie along polarity inversion lines, measurement of their tilt angle help us to understand the distribution of the solar  magnetic field across the solar disk. In contrast to sunspots, filaments are distributed at all latitudes, so the long term variation of the distribution of poloidal and toroidal components of magnetic fields may be better studied if we follow the filament distribution.  

\end{itemize}
 In this paper we announce the availability of the longest digitized archive of $H_{\alpha}$ dataset (1914 -- 2007) from the Kodaikanal Solar Observatory. The data will be made public through its portal at \url{http://kso.iiap. res.in/data}. We also show representative results from this archive to prove the potential of such an archive which demands further detailed investigations. In our subsequent studies on filaments, we want to combine multi-wavelength analysis of the historical data from KSO as hinted in this paper. This can lead to the classification of filaments into active region filaments, quiescent filaments. We also want to explore the hemispheric differences in filament behaviours and find more proxies to predict the future cycles. As a next step, we will employ different techniques for getting rid of filament fragmentation to estimate the filament lengths correctly. The length determination may prove to be vital in terms of its correlation to different statistical studies related to filament eruption and CMEs \citep{{2011JGRA..116.4220R},{2008AnGeo..26.3025F}}. We hope that this KSO archive and some of the new results as presented here  will provide new momentum for the long term study of H-alpha. 
  \section{acknowledgements}
 We would like to thank all the observers at Kodaikanal over 100 years for their contribution to build this enormous resource. The current high resolution digitisation process was initiated by Prof. Jagdev singh and we thank him for his important  contribution to the project.This data is now available for public use at \url{http://kso.iiap. res.in/data}. We also thank the Science \& Engineering  Research Board (SERB) for the project grant (EMR/2014/000626).


\begin{thebibliography}{}
\expandafter\ifx\csname natexlab\endcsname\relax\def\natexlab#1{#1}\fi

\bibitem[{{Ananthakrishnan}(1952)}]{1952Natur.170..156A}
{Ananthakrishnan}, R. 1952, \nat, 170, 156

\bibitem[{Chatterjee {et~al.}(2016)Chatterjee, Banerjee, \&
  Ravindra}]{0004-637X-827-1-87}
Chatterjee, S., Banerjee, D., \& Ravindra, B. 2016, The Astrophysical Journal,
  827, 87

\bibitem[{{Chen} {et~al.}(2008){Chen}, {Innes}, \&
  {Solanki}}]{2008A&A...484..487C}
{Chen}, P.~F., {Innes}, D.~E., \& {Solanki}, S.~K. 2008, \aap, 484, 487

\bibitem[{{Evershed}(1907)}]{1907MNRAS..67..477E}
{Evershed}, J. 1907, \mnras, 67, 477

\bibitem[{{Evershed}(1908)}]{1908MNRAS..68..515E}
---. 1908, \mnras, 68, 515

\bibitem[{{Filippov} \& {Koutchmy}(2008)}]{2008AnGeo..26.3025F}
{Filippov}, B., \& {Koutchmy}, S. 2008, Annales Geophysicae, 26, 3025

\bibitem[{Gilbert {et~al.}(2000)Gilbert, Holzer, Burkepile, \&
  Hundhausen}]{0004-637X-537-1-503}
Gilbert, H.~R., Holzer, T.~E., Burkepile, J.~T., \& Hundhausen, A.~J. 2000, The
  Astrophysical Journal, 537, 503

\bibitem[{{Gopalswamy} {et~al.}(2000){Gopalswamy}, {Hanaoka}, \&
  {Hudson}}]{2000AdSpR..25.1851G}
{Gopalswamy}, N., {Hanaoka}, Y., \& {Hudson}, H.~S. 2000, Advances in Space
  Research, 25, 1851

\bibitem[{Gopalswamy {et~al.}(2003)Gopalswamy, Shimojo, Lu, Yashiro, Shibasaki,
  \& Howard}]{0004-637X-586-1-562}
Gopalswamy, N., Shimojo, M., Lu, W., {et~al.} 2003, The Astrophysical Journal,
  586, 562

\bibitem[{{Hao} {et~al.}(2015){Hao}, {Fang}, {Cao}, \&
  {Chen}}]{2015ApJS..221...33H}
{Hao}, Q., {Fang}, C., {Cao}, W., \& {Chen}, P.~F. 2015, \apjs, 221, 33

\bibitem[{Li(2010)}]{doi:10.1111/j.1365-2966.2010.16508.x}
Li, K.~J. 2010, Monthly Notices of the Royal Astronomical Society, 405, 1040

\bibitem[{Li {et~al.}(2007)Li, Li, Gao, Mu, Chen, \& Su}]{Li2007}
Li, K.~J., Li, Q.~X., Gao, P.~X., {et~al.} 2007, Journal of Astrophysics and
  Astronomy, 28, 147

\bibitem[{Low(1982)}]{ROG:ROG841}
Low, B.~C. 1982, Reviews of Geophysics, 20, 145

\bibitem[{{Makarov} \& {Sivaraman}(1983)}]{1983SoPh...85..227M}
{Makarov}, V.~I., \& {Sivaraman}, K.~R. 1983, \solphys, 85, 227

\bibitem[{Martin(1998)}]{Martin1998}
Martin, S.~F. 1998, Solar Physics, 182, 107

\bibitem[{{McIntosh}(1972)}]{1972RvGSP..10..837M}
{McIntosh}, P.~S. 1972, Reviews of Geophysics and Space Physics, 10, 837

\bibitem[{{Mouradian} \& {Soru-Escaut}(1994)}]{1994A&A...290..279M}
{Mouradian}, Z., \& {Soru-Escaut}, I. 1994, \aap, 290, 279

\bibitem[{{Rausaria} {et~al.}(1993{\natexlab{a}}){Rausaria}, {Gupta},
  {Selvendran}, {Sundara Raman}, \& {Singh}}]{1993SoPh..146..259R}
{Rausaria}, R.~R., {Gupta}, S.~S., {Selvendran}, R., {Sundara Raman}, K., \&
  {Singh}, J. 1993{\natexlab{a}}, \solphys, 146, 259

\bibitem[{{Rausaria} {et~al.}(1993{\natexlab{b}}){Rausaria}, {Sundara Raman},
  {Aleem}, \& {Singh}}]{1993SoPh..146..137R}
{Rausaria}, R.~R., {Sundara Raman}, K., {Aleem}, P.~S.~M., \& {Singh}, J.
  1993{\natexlab{b}}, \solphys, 146, 137

\bibitem[{{Ruzmaikin} {et~al.}(2011){Ruzmaikin}, {Feynman}, \&
  {Stoev}}]{2011JGRA..116.4220R}
{Ruzmaikin}, A., {Feynman}, J., \& {Stoev}, S.~A. 2011, Journal of Geophysical
  Research (Space Physics), 116, A04220

\bibitem[{{Sheeley} {et~al.}(2011){Sheeley}, {Cooper}, \&
  {Anderson}}]{2011ApJ...730...51S}
{Sheeley}, Jr., N.~R., {Cooper}, T.~J., \& {Anderson}, J.~R.~L. 2011, \apj,
  730, 51

\bibitem[{Sonka {et~al.}(2014)Sonka, Hlavac, \& Boyle}]{sonka2014image}
Sonka, M., Hlavac, V., \& Boyle, R. 2014, Image Processing, Analysis, and
  Machine Vision (Cengage Learning)

\bibitem[{{Sundara Raman} {et~al.}(1994){Sundara Raman}, {Aleem}, {Singh},
  {Selvendran}, \& {Thiagarajan}}]{1994SoPh..149..119S}
{Sundara Raman}, K., {Aleem}, S.~M., {Singh}, J., {Selvendran}, R., \&
  {Thiagarajan}, R. 1994, \solphys, 149, 119

\bibitem[{{Sundara Raman} {et~al.}(2001){Sundara Raman}, {Ramesh}, \&
  {Selvendran}}]{2001BASI...29...77S}
{Sundara Raman}, K., {Ramesh}, K.~B., \& {Selvendran}, R. 2001, Bulletin of the
  Astronomical Society of India, 29, 77

\bibitem[{{Tlatov} {et~al.}(2016{\natexlab{a}}){Tlatov}, {Kuzanyan}, \&
  {Vasil'yeva}}]{2016ASPC..504..241T}
{Tlatov}, A.~G., {Kuzanyan}, K.~M., \& {Vasil'yeva}, V.~V. 2016{\natexlab{a}},
  in Astronomical Society of the Pacific Conference Series, Vol. 504, Coimbra
  Solar Physics Meeting: Ground-based Solar Observations in the Space
  Instrumentation Era, ed. I.~{Dorotovic}, C.~E. {Fischer}, \& M.~{Temmer}, 241

\bibitem[{{Tlatov} {et~al.}(2016{\natexlab{b}}){Tlatov}, {Kuzanyan}, \&
  {Vasil'yeva}}]{2016SoPh..291.1115T}
{Tlatov}, A.~G., {Kuzanyan}, K.~M., \& {Vasil'yeva}, V.~V. 2016{\natexlab{b}},
  \solphys, 291, 1115

\bibitem[{{Zhang} {et~al.}(2012){Zhang}, {Chen}, {Xia}, \&
  {Keppens}}]{2012A&A...542A..52Z}
{Zhang}, Q.~M., {Chen}, P.~F., {Xia}, C., \& {Keppens}, R. 2012, \aap, 542, A52

\bibitem[{Zharkova {et~al.}(2005)Zharkova, Ipson, Benkhalil, \& Zharkov}]{zhar}
Zharkova, V., Ipson, S., Benkhalil, A., \& Zharkov, S. 2005, Artificial
  Intelligence Review, 23, 209

\bibitem[{Zharkova \& Schetinin(2003)}]{Zharkova2003}
Zharkova, V.~V., \& Schetinin, V. 2003, A Neural-Network Technique for
  Recognition of Filaments in Solar Images, ed. V.~Palade, R.~J. Howlett, \&
  L.~Jain (Berlin, Heidelberg: Springer Berlin Heidelberg), 148--154

\end{thebibliography}

 \end{document}